\documentclass{PoS}

\title{Review of top charge asymmetries for LHC run II}

\ShortTitle{Top charge asymmetries for LHC}

\author{Susanne Westhoff\\
        Institute for Theoretical Physics, Heidelberg University, 69120 Heidelberg, Germany\\
        E-mail: \email{westhoff@thphys.uni-heidelberg.de}}

\abstract{This is a brief review of charge asymmetries in top-antitop quark production at the LHC. Proposed asymmetry observables in processes of associated top-pair production $t\bar t$ + jet, $t\bar t + W$, and $t\bar t + \gamma$, are analyzed and compared. While the focus is on standard-model predictions, we also comment on the sensitivity to possible new physics and observation prospects during the second runtime of the LHC.}

\FullConference{9th International Workshop on the CKM Unitarity Triangle\\
		28  November - 3 December 2016\\
		Tata Institute for Fundamental Research (TIFR), Mumbai, India}

\begin{document}

\section{Introduction}
\noindent The precise predictions and measurements of abundantly produced top-quarks at the LHC allow us to probe features beyond cross sections, such as the charge asymmetry in top-antitop production. In the standard model (SM), the top charge asymmetry provides us with an important test of QCD. In this respect, the charge asymmetry is complementary to charge-symmetric observables, as it probes different structures of QCD contributing to the hard production process. Beyond the standard model, the charge asymmetry can reveal potential new particles with top-quark interactions with a high sensitivity, provided that a good experimental accuracy can be achieved.

In inclusive top-pair production from proton-proton collisions at the LHC, the observation of a charge asymmetry is hindered by the huge background from symmetric $gg\to t\bar t$ parton interactions. Furthermore, in the forward-backward symmetric experiment, the charge asymmetry at the parton level is not translated into an angular forward-backward asymmetry at the hadron level. Therefore, rapidity distributions of top and antitop quarks can only give limited access to the original charge asymmetry in the hard scattering process. To overcome these complications, two strategies have been pursued to find asymmetry observables that are tailored to the LHC environment: to design observables with a \emph{suppressed gluon-gluon background}; and to exploit final-state kinematics that give the most \emph{direct access to the charge asymmetry} in the partonic process. This approach has led to considering processes of associated top-antitop production, namely $t\bar t$ + jet, $t\bar t + W$, and $t\bar t +\gamma$, where the additional particle in the final state is used to define new charge asymmetry observables.

In this write-up, we review three proposed observables of the charge asymmetry in associated top-pair production. Our goal is to provide the reader with a concise reference of available asymmetry predictions for the LHC. We compare these observables with respect to their sensitivity to physics in and beyond the standard model. Finally, we briefly comment on proposals to observe the charge asymmetry in inclusive top-antitop production in the boosted regime at ATLAS and CMS, and in the forward region with the LHCb detector.

\section{Top energy asymmetry in $t\bar t$ + jet production}
\noindent In top-antitop production with an associated hard jet, the charge asymmetry can be probed through the difference between top and antitop energies, $\Delta E = E_t - E_{\bar t}$, in the partonic center-of-mass system. The energy asymmetry is defined by~\cite{Berge:2013xsa}
\begin{equation}
A_E = \frac{\sigma_{t\bar t j}(\Delta E > 0) - \sigma_{t\bar t j}(\Delta E < 0)}{\sigma_{t\bar t j}(\Delta E > 0) + \sigma_{t\bar t j}(\Delta E < 0)}.
\end{equation}
This observable probes the charge asymmetry in partonic quark-gluon interactions. Due to the higher abundance of quark-gluon versus quark-antiquark states in proton-proton interactions, the gluon-gluon background is significantly reduced. This is especially true if the jet is energetic and emitted in the direction of the incoming quark. A strong kinematic cut on the transverse momentum of the hardest jet, $p_T(j)$, can thus be used to efficiently suppress the gluon-gluon background.

In QCD, the energy asymmetry in $t\bar t$ + jet production is induced at the leading order (LO), where it corresponds with an angular asymmetry of the (quark-)jet in the top-antitop center-of-mass frame. Measuring the energy distributions of the top and antitop quarks thus gives direct access to the charge asymmetry at parton level. Beyond LO, the energy asymmetry can be affected by additional jet radiation in the production process and from top decays. In Ref.~\cite{Berge:2016oji}, it has been shown that the observable is stable under next-to-leading order (NLO) QCD corrections to top production and that effects of radiation in top decay are expected to be small, if the jet is sufficiently hard and emitted perpendicular to the beam axis in the parton center-of-mass frame. In this regime, with suitable phase-space cuts, the energy asymmetry is predicted to reach $A_E = -6.5^{+0.1}_{-0.2}\%$ at $\sqrt{s}=13$ TeV, corresponding with a cross section of $\sigma_{t\bar t j} = 1$ pb. The quoted uncertainty is due to the factorization and renormalization scale dependence.

Beyond the SM, the energy asymmetry is sensitive to the same kind of physics leading to a modified rapidity asymmetry in inclusive top-pair production. However, it probes new contributions in a different kinematic regime, which opens the possibility to observe new physics in the energy asymmetry that are hidden from observables in inclusive top-pair production. A detailed study of the sensitivity of the energy asymmetry to axigluons has been performed in Ref.~\cite{Alte:2014toa}.

\section{Top rapidity asymmetry in $t\bar t + W$ production}
\noindent In $t\bar t + W$ production, the presence of the additional $W$ boson strongly affects the production process of top-quarks and thereby the top charge asymmetry~\cite{Maltoni:2014zpa}. As in inclusive top-pair production, a rapidity asymmetry can be defined as
\begin{equation}
A_{\eta}^W = \frac{\sigma_{t\bar t W}(\Delta |\eta| > 0) - \sigma_{t\bar t W}(\Delta |\eta| < 0)}{\sigma_{t\bar t W}(\Delta |\eta| > 0) + \sigma_{t\bar t W}(\Delta |\eta| < 0)},
\end{equation}
with $\Delta |\eta| = |\eta_t| - |\eta_{\bar t}|$, where $\eta_t$ and $\eta_{\bar t}$ denote the pseudo-rapidities of the top and antitop quarks. Since the $W$ boson in $t\bar t + W$ production can be emitted only from a light quark, gluon-gluon contributions are absent up to the NNLO in QCD. In the SM, the top-antitop charge asymmetry is induced first at the NLO in quark-antiquark interactions. At the LHC with $\sqrt{s}=13$ TeV, the predicted rapidity asymmetry at NLO is $A_{\eta}^W = -2.24^{+0.43}_{-0.32}\%$. The result includes a parton shower, and uncertainties are due to scale variations.

The emission of the $W$ boson from the initial-state quark implies the production of polarized top-quarks. This polarization imprints itself on the angular distributions of the top decay pro\-ducts from $t\to b\ell^+$ and thus generates different rapidity distributions for bottom-quarks and leptons from top and antitop decays at the LO. The rapidity asymmetry $A_{\eta}^W$ with $\Delta |\eta| = \eta_{b,\ell^+} - \eta_{\bar{b},\ell^-}$ can therefore be used to probe the top-quark polarization in the production process. While a measurement of $A_{\eta}^W$ for reconstructed top-quarks at the LHC requires a large data set, comparably better statistical precision can be obtained for the lepton and bottom-quark asymmetries, which are larger in magnitude.

Beyond the SM, the rapidity asymmetries in $t\bar t + W$ production can help to observe and disentangle potential contributions from new physics, especially if they affect the polarization of the top-quarks. As for the energy asymmetry, modifications of the charge asymmetry are generally expected to be larger than in inclusive top-pair production, mainly because $A_{\eta}^W$ does not suffer from gluon-gluon background.

\section{Top rapidity asymmetry in $t\bar t + \gamma$ production}
\noindent A third option to observe the top charge asymmetry is to consider the rapidity asymmetry in $t\bar t + \gamma$ production~\cite{Aguilar-Saavedra:2014vta},
\begin{equation}
A_y^\gamma = \frac{\sigma_{t\bar t\gamma}(\Delta |y| > 0) - \sigma_{t\bar t\gamma}(\Delta |y| < 0)}{\sigma_{t\bar t\gamma}(\Delta |y| > 0) + \sigma_{t\bar t\gamma}(\Delta |y| < 0)},
\end{equation}
with $\Delta |y| = |y_t| - |y_{\bar t}|$, where $y_t$ and $y_{\bar t}$ are the rapidities of the top and antitop quarks. While the source of the QCD charge asymmetry is the same as in inclusive top-pair production, the presence of the photon enhances the fraction of quark-antiquark versus gluon-gluon interactions. This effect lead to a larger net rapidity asymmetry, due to a reduced gluon-gluon background.

The main challenge in measuring $A_y^\gamma$ consists in suppressing photon radiation from top decays, which makes up the dominant part of the cross section, but does not contribute to the charge asymmetry. This can be achieved by applying selection cuts on the photon kinematics and a veto on radiative top and $W$ boson decays. After cuts, the rapidity asymmetry in $t\bar t + \gamma$ production at the LO in QCD is predicted to be $A_y^\gamma = -3.5\pm 0.2\%$ at $\sqrt{s} = 14$ TeV. Quoted uncertainties are due to limited statistics in Monte-Carlo simulation. A crucial limiting factor for an observation of the rapidity asymmetry in $t\bar t + \gamma$ production will be the experimental statistical sensitivity, due to the small production cross section.

The fact that the photon re-weights the contributions to the asymmetry from initial up- versus down-quarks and from initial quarks versus gluons can be used to search for new physics. Speci\-fi\-cally, charge-asymmetric contributions that cancel in the rapidity asymmetry in inclusive top-pair production can be visible in $A_y^\gamma$.

\section{Boosted top asymmetry and lepton charge asymmetry at LHCb in inclusive $t\bar t$ production}
\noindent An alternative to associated top-antitop production is to consider the rapidity asymmetry in inclusive top-pair production in a kinematic regime with enhanced quark-antiquark contributions~\cite{Kuhn:2011ri}. It has been shown that the asymmetry can be enhanced up to a few percent in regions of high invariant mass $M_{t\bar t}$ or high boost $\beta_{t\bar t}$ of the top-antitop pair~\cite{Bernreuther:2012sx}. The main challenge to observe the rapidity asymmetry in these regions consists in suppressing QCD background close to the beam axis and/or dealing with a strong reduction of data statistics.

In this spirit, a measurement of the top charge asymmetry in the forward region at LHCb has been proposed~\cite{Kagan:2011yx}. While one of the top-quarks escapes the detector, the top-antitop charge asymmetry can still be observed through a charge asymmetry of the leptons from top or antitop decays, where the lepton charge reflects the top charge. A severe complication for a measurement of this lepton charge asymmetry is the control of large background from mis-tagged $W$ + jet, $Z$ + jet, or single-top events~\cite{Gauld:2014pxa}.

\section{Summary}
\noindent The main features of the discussed charge asymmetry observables in associated top-antitop production are summarized in Table~\ref{tab:asymmetries}.
\begin{table}[!tb]
\centering
\renewcommand{\arraystretch}{1.5}
\begin{tabular}{|c|c|c|c|}
\hline
 & $A_E$ in $t\bar t$ + jet & $A_{\eta}^W$ in $t\bar t + W$ & $A_{y}^\gamma$ in $t\bar t + \gamma$\\
\hline
production cross section & $\mathcal{O}(500)$ pb & $\mathcal{O}(700)$ fb & $\mathcal{O}(100)$ fb\\
charge asymmetry in QCD & LO & NLO & LO\\
available prediction in QCD & NLO & NLO & LO\\ 
asymmetry in parton channel & $qg$ & $q\bar q$ & $q\bar q$\\
gluon-gluon background & reduced & absent up to NNLO & reduced\\
\hline
\end{tabular}
\caption{Comparison of charge asymmetry observables in associated top-antitop production at the LHC with $\sqrt{s}=13$ TeV.}
\label{tab:asymmetries}
\end{table}
 All three observables offer access to the charge asymmetry in a way that is complementary to inclusive top-antitop production and to the respective other associated channels. The rapidity asymmetry in $t\bar t + W$ production offers the best suppression of the gluon-gluon background. In turn, at the parton level, the energy asymmetry in $t\bar t$ + jet and the rapidity asymmetry in $t\bar t + \gamma$ production are relatively larger, since they occur at the LO in QCD. Predictions beyond the leading contribution in QCD are currently only available for the energy asymmetry, which has been calculated at the NLO.
 
The observation prospects of the three asymmetries at the LHC strongly depend on experimental aspects, which makes their comparison difficult. The energy asymmetry in $t\bar t$ + jet is certainly based on the largest data set, such that kinematic cuts are a viable option to enhance the asymmetry by focusing on specific phase-space regions. Due to the smaller cross section for $t\bar t + W$ and $t\bar t + \gamma$ production, kinematic cuts are less suitable here. In turn, the $W$ boson or photon might provide a cleaner signature than an additional hard jet. From the theory point of view, all three observables have interesting and complementary features that are worthwhile exploring. We therefore strongly suggest to pursue their measurement in the increasingly large data set obtained from the second runtime of the LHC.

\section{Acknowledgements}
\noindent A warm acknowledgement goes to the organizers of CKM 2017 for an interesting conference and for introducing us visitors to the character of Mumbai.
SW acknowledges funding by the Carl Zeiss foundation through a \emph{Junior-Stiftungsprofessur}.

\end{document}